\documentclass[12pt,preprint]{aastex}
\begin{document}

\newcommand{\kms}{km\ s$^{-1}$}
\newcommand{\cc}{cm$^{-3}$}
\newcommand{\co}{$^{12}$CO}
\newcommand{\thco}{$^{13}$CO}
\newcommand{\ceo}{C$^{18}$O}
\newcommand{\hco}{HCO$^+$}
\newcommand{\hthco}{H$^{13}$CO$^+$}
\newcommand{\cts}{C$^{34}$S}
\newcommand{\nh}{N$_2$H$^+$}
\newcommand{\vchan}{$v_{chan}$}
\newcommand{\vout}{$v_{out}$}
\newcommand{\venv}{$v_{env}$}

\title{Complex Molecules in the L1157 Molecular Outflow}
\author{H\'ector G.~Arce}
\affil{Department of Astronomy,
Yale University, P.O.~Box 208101, New Haven, CT 06520-8101} 
\email{hector.arce@yale.edu} 
\author{Joaqu\'{i}n Santiago-Garc\'{i}a}
\affil{Observatorio Astron\'omico Nacional, Alfonso XII 3, E-28014 Madrid, Spain}
\email{j.santiago@oan.es}
\author{Jes K. J{\o}rgensen}
\affil{Argelander-Institut f\"ur Astronomie, University of Bonn, Auf dem H\"ugel 71, 53121 Bonn, Germany}
\email{jes@astro.uni-bonn.de}
\author{Mario Tafalla}
\affil{Observatorio Astron\'omico Nacional, Alfonso XII 3, E-28014 Madrid, Spain}
\email{m.tafalla@oan.es}
\and
\author{Rafael Bachiller}
\affil{Observatorio Astron\'omico Nacional, Alfonso XII 3, E-28014 Madrid, Spain}
\email{r.bachiller@oan.es}

\shorttitle{Complex Molecules in L1157}
\shortauthors{Arce et al.}

\begin{abstract}
We report the detection of complex organic molecules in the young protostellar outflow L1157.
We identify lines from HCOOCH$_3$, CH$_3$CN, HCOOH and C$_2$H$_5$OH
at the position of the B1 shock in the blueshifted lobe,
making it the first time that complex 
species have been detected towards
a molecular outflow powered by a young low-mass protostar.  
The time scales associated with
the warm outflow gas ($< 2 \times 10^3$ yr)
are too short for the complex molecules to have 
formed in the gas phase after the
shock-induced sputtering of the grain mantles. 
It is more likely that the complex species 
formed in the surface of grains and were then ejected from
 the grain mantles by the shock.
The formation of complex molecules in the grains 
of low-mass star forming regions must
be relatively efficient, and our results show the importance of 
considering the impact of 
outflows 
when studying complex molecules around protostars.
The relative abundance with respect to methanol
of most of the detected complex molecules is similar
 to that of hot cores and molecular clouds
in the galactic center region, which suggests that 
the mantle composition of the dust in the 
L1157 dark cloud is similar to dust in those regions. 

\end{abstract}


\section{Introduction}
\label{intro}

Shocks from protostellar outflows
 heat and compress the surrounding medium thereby triggering
different  processes (e.g., grain
disruption, ice gain mantle sublimation, etc.) that can release
molecules trapped in the grains into the gas phase. In this way, 
outflows contribute to the chemical enrichment 
of the gaseous environment surrounding young stars.
This is 
clearly demonstrated by multi-molecular line
observations  reflecting enhanced
molecular abundances either in shocks associated with the outflows or
in the outflow cavity walls in the protostellar envelopes
(e.g., Garay et al.~1998; Bachiller et al.~2001; 
Arce \& Sargent 2006; J\o rgensen et al.~2007).
In most cases these
molecules have similar velocity and structure as the CO 
lines that trace the outflows, and
it is clear that the outflow is responsible for their overabundance.

The limited outflow chemistry studies show shock-triggered
overabundance of parent (or ``first generation'') 
species ---molecules that are released
directly into the gas phase from the icy dust mantles, like H$_2$CO
and CH$_3$OH--- as well as other simple molecules 
presumably formed in the warm gas (e.g., SO, HCO$^+$).
Chemical models indicate that larger organic
molecules (made of 7 atoms or more) can form on grain 
surfaces (e.g., Garrod \& Herbst 2006) 
and should be present in the gas phase along with the other
simpler molecules observed near outflow shocks
---as suggested by Chandler et al.~(2005) 
and Remijan \& Hollis (2006).
Yet, up to date, conclusive evidence for
shock-triggered overabundances 
in molecular outflows from low-mass stars ($M < 2M_{\sun}$)
exists only
for species with six atoms or less.  
Here we report the first detection of HCOOH, CH$_3$CN, HCOOCH$_3$, C$_2$H$_5$OH
 in the the L1157 molecular outflow.

The L1157 dark cloud (at a distance
of about 440 pc) harbors  
an embedded low-mass stellar object, 
L1157-mm (IRAS 20386+6751), that powers a young
molecular outflow (Umemoto et al.~1992). 
The powerful L1157 outflow
is comprised of several shocks seen as bright knots
in near and mid infrared images 
(Davis \& Eisloeffel 1995; Cabrit et al.~1998;
Looney et al.~2007).
Millimeter observations
of many molecular emission lines
show that L1157 is a chemically rich outflow exhibiting an
overabundance of SiO, CH$_3$OH, H$_2$CO, CN, HCN, SO,
and a number of other species by a factor
of a few tens to more than $10^5$,
depending on the molecule (Bachiller \& P\'erez Guiti\'errez
1997, hereafter BP97).  
In the blue (southern) lobe, 
the emission from these species  is
brightest towards two to three clumps  (Bachiller et al.~2001),
and  high-resolution interferometer observations 
reveal the clumps are associated with shell-like structures created
by two jet-driven bow shocks (Zhang et al.~1995; Gueth et al.~1996). 
These results indicate that the observed 
chemical enhancements are induced by the outflow shocks.

\section{Observations and Results}
\label{obs}

Observations of the lines shown in Table 1 were obtained using the IRAM 30~m telescope 
in Pico Veleta, Spain, in June 2007. 
Our observations concentrated in the brightest CO clump in the
blue lobe of the L1157 outflow (i.e., the B1 clump), located 
at an offset of  $\Delta \alpha = 22.5\arcsec$,  $\Delta \delta = -64.5\arcsec$ from the 
L1157-mm protostar at $\alpha$(J2000) = $20^h39^m06^s.2$,  
$\delta$(J2000) = $68\arcdeg02\arcmin16\arcsec$ (Bachiller et al.~2001). 
The B1 clump is far enough from the source and the telescope
beam is sufficiently small (see below) that 
any molecular line emission associated with the gas immediately
surrounding L1157-mm (i.e., within $10^4$ AU) does not contaminate our observations.
Two different spectral configurations
were used in order to include as many line transitions as possible of  
HCOOH, C$_2$H$_5$CN, CH$_3$CN, and HCOOCH$_3$ (A and E) ---complex molecules that
have been detected in the immediate surroundings
of deeply embedded low-mass protostars
 (Cazaux et al.~2003; Bottinelli et al.~2007).
Both spectral configurations included four receivers, each simultaneously connected 
to a 256 channel filter bank with a spectral resolution of 1 MHz and a unit of an
autocorrelator,  that provided simultaneous observations
 at four different frequencies ranging from 90.1 to 257.6 GHz.  The autocorrelator was set to
 provide spectral resolutions of 40 and 320 kHz and bandwidths between 40 and 320 MHz. 
 All observations were obtained in wobbler switching mode with a $90\arcsec$
 throw ---large enough to avoid any possible ``contaminating'' emission from the 
rest of the blueshifted lobe or the immediate surroundings of L1157-mm. 
Verification of the telescope pointing  approximately every two hours, 
 indicated that the pointing accuracy was between 2 and $4\arcsec$.
A total integration time of four to ten hours, 
and average system temperatures between 120 K and
830 K, depending on the frequency,
resulted in spectra with rms of 0.002 to 0.01 K.
The telescope beamwidth (FWHM)  at 
90.2, 98.6, 135.7, 226.7, and 257.5 GHz is $27\arcsec$, 
$25\arcsec$, $18\arcsec$, $11\arcsec$, and $10\arcsec$, respectively.
Spectral data presented here are in units of
main beam brightness temperature, $T_{mb}$ 
(see Rohlfs \& Wilson 2000). 

We detect, at a 3 $\sigma$ level or higher,  five HCOOCH$_3$-A transitions, 
 four HCOOCH$_3$-E transitions, three CH$_3$CN transitions, two HCOOH transitions and one C$_2$H$_5$OH
 transition (see Table 1). 
These were identified using the CDMS and JPL catalogs (M\"uller et al.~2001; Pickett et al.~1998).  
We are confident that the emission arises from the L1157 outflow's 
 blue lobe as all lines are blueshifted by 1.5 to 4.0 \kms \/  with 
respect to the L1157 cloud central velocity ($V_{LSR} = 2.7$ \kms)  
and have a velocity width between 2 and 7~\kms \/
as expected from the velocity distribution
of other species associated with the L1157 outflow, observed towards
  B1 by BP97. Figure 1 shows an example of our detect lines. 
 We used the rotational diagram method 
to estimate the rotational temperature ($T_{rot}$)
and column densities  
of the species for which we detect two or more transitions
(see, e.g., Bisschop et al.~2007).
We only detect one transition of C$_2$H$_5$OH, so we estimate 
its column density assuming $T_{rot} \sim 25$~K (similar to that of
HCOOCH$_3$)
and follow Requena-Torres et al.~(2006). For all lines we assume the 
emission is optically thin and the gas is in local thermodynamic 
equilibrium. 
In Table 2 we show our results with a correction for beam dilution,
assuming  a source size of 10.5\arcsec .  
We base our assumption of the source 
size on the high-resolution interferometer 
 CH$_3$OH observations of the L1157 blueshfited lobe by 
Benedettini et al.~(2007) which show that 
 B1 is mostly made of two subclumps. Taken together, the 
emission from these two 
 subclumps extends about 10.5\arcsec \/ 
(estimated from the geometric mean of the minor and major axis of the emission,
 see Figure 2 of Benedettini et al.~2007). 
Whether the complex molecules are formed as a by product of methanol in the
warm gas or are ejected from the grain mantle with the
 methanol (see below), 
 we expect the emission from the observed 
complex molecules to
be coincident with that of CH$_3$OH (see also, Liu et al.~2001). The CH$_3$CN emission 
is not corrected for source
size as the beam size for all three transitions of this molecule is smaller 
than 10.5\arcsec.

Our estimates indicate that most of the detected molecules have
$T_{rot}$ between 10 and 30~K. This 
is lower than the estimated gas 
temperature of B1 ($T \sim 80 - 300$ K, see below), 
as well as significantly lower  
than the temperature needed to evaporate
off most of the species present in icy grain mantles
(e.g., van Dishoeck \& Blake 1998). Even so, similar
low temperatures have also been derived for the 
same complex molecules in the gas immediately 
surrounding
low-mass protostars (Cazaux et al.~2003; Bottinelli et al. 2004; 2007).
It is possible that the low estimates
of $T_{rot}$ are due to our poorly constrained rotational diagrams,
as we only detect two to four transitions (mostly at low energies) for
each molecule. In addition,  non-LTE effects may 
influence the derived temperature from the rotational diagram
(Goldsmith \& Langer 1999). 
Shock-heated molecular gas takes only a few hundred years to 
cool from 100 K to 20 K (Bergin et al.~1998), and it is plausible  that 
some of the detected molecules come from gas that has 
cooled down to the estimated low temperatures. 
Further observations are needed to
distinguish with certainty which
of the scenarios suggested above affect our estimates. 

We calculate the abundance of each molecule
($X = N/N_{H_2}$) using our estimate of the 
molecule's column density ($N$) and the estimate of H$_2$ column 
density ($N_{H_2}$) towards B1 obtained by BP97. The latter was
obtained from single-dish observations of high-velocity CO
outflow emission and assuming a standard CO/H$_2$ ratio of 
$10^{-4}$. Uncertainties on the size of B1, the CO/H$_2$ ratio and
the opacity of the CO line introduce uncertainties on our estimate
of the absolute value of the abundance. However, these 
uncertainties are less important when comparing the relative
abundance between the different molecules.  
 
\section{Discussion}
\label{disc}
Our results clearly indicate that complex organic molecules
can be found in outflows from low-mass stars. 
The interesting question that follows is, how 
are these complex molecules formed? The two 
main competing scenarios propose that
complex organic molecules mainly 
form either in the warm gas phase or on grain surfaces.
The gas phase models were originally conceived to
explain the chemical richness of hot cores. 
In these objects  
the radiation from a massive protostar is believed to heat 
the dense ($n \sim 10^6 - 10^7$~cm$^{-3}$) inner core to 
temperatures higher than 100~K
where the icy grain mantles evaporate injecting molecules
like H$_2$O, H$_2$CO, CH$_3$OH into the
gas-phase (e.g., Kurtz et al.~2000). The gas phase models indicate that
subsequent chemical ion-molecule reactions in the warm gas can result
in the formation of other, more complex, molecules, such as
HCOOCH$_3$ and C$_2$H$_5$OH (the so called daughter or
``second generation'' species) (Charnley et al.~1992; Caselli et
al.~1993). Alternatively, in the grain surface model,
the observed complex organic molecules are formed
on the grain surface and are later
released into the gas phase 
(with H$_2$CO and CH$_3$OH)
when the grain mantles evaporate (e.g., Hasegawa et al.~1992;
Hasegawa \& Herbst 1993).

In both models, a heat source is necessary to evaporate
the icy grain mantle. 
At the position of B1, the outflow shock
is the only source that can trigger the ejection of molecules
from the grain surface.
Using the known characteristics
of the
 L1157 outflow shocks, we can deduce which of the two mechanisms
is most likely responsible for the formation 
of the detected complex molecules. 
Multi-transition millimeter observations of NH$_3$, CO and SiO
indicate that the gas associated with B1 has a
temperature between 80 and 300 K  and a
density of about $3 \times 10^5$ cm$^{-3}$  
(Tafalla \& Bachiller 1995;
Umemoto et al.~1999; Hirano \& Taniguchi 2001;
 Nisini et al.~2007). 
The shocked region associated with B1 extends
no more than 20\arcsec \/ along the outflow
axis (see H$_2$ image in Zhang et al.~2000),
and Nisini et al.~(2007) indicate that 
shocks in L1157 have velocities larger 
than $\sim 30$~\kms . We, therefore, estimate
the time it took the shock to transverse the area
associated with B1 to be no more than about
1400 yrs. Molecular gas heated by a
shock with a velocity of about 30~\kms \/
takes a few hundred years to cool down below 
100 K (Bergin et al.~1998). Hence, the  
gas in the B1 region has been above 100~K
for no more than 2000 yrs.

The B1 region is hot enough for grain
mantles to evaporate and it is dense enough for 
gas-phase chemical
reactions that result in complex molecules
to occur (e.g., Millar et al.~1991). However, 
gas-phase models predict
the maximum abundance of complex molecules 
in the gas to occur 
between a few $10^4$ yrs and a few $10^5$ yrs after
the parent molecules are released into the gas phase (see Millar et al.~1991) 
---considerably longer than the timescale estimated above---
and the maximum abundance of HCOOCH$_3$ 
predicted by these models 
is a factor of ten less
than our observed abundance. Moreover, recent results
by Horn et al.~(2004) indicate that gas-phase
production of HCOOCH$_3$ is much less efficient
than previously considered by gas-phase models.
It is highly improbable that most of the observed
complex molecules in the L1157 outflow are
produced in the gas phase. 
We therefore conclude that
the relatively high abundance of complex molecules in the
L1157 outflow
is better explained by the formation of these
species on the grain surface and  their subsequent 
release into the gas phase caused by the outflow shock. 
We note that our results do not discard the possibility
of small differences in the formation mechanism of
these molecules, and that a small fraction of 
some of the species may exist in the gas phase 
independent of outflow shocks.
For example, HCOOH has been
observed in a quiescent dark cloud, 
yet HCOOCH$_3$ has only been observed in active regions
 (Turner et al.~1999; Requena-Torres et al.~2007).
The derived HCOOH abundance in these studies is
about $10^{-10}$, two orders of magnitude smaller
than our abundance estimates in L1157. 
Hence, only a negligible fraction of the 
complex species observed in L1157 could be due to processes 
unrelated to the outflow shock.

Recent surveys of abundances of complex species in hot cores,
hot corinos, 
(the presumed low-mass
counterparts of hot cores), 
and molecular clouds in the 
galactic center (GC) region favor
the grain surface formation scenario 
(e.g., Bottinelli et al. 2007; Bisschop et al. 2007; Requena-Torres et al. 2006; 2007).
The abundance ratio of the complex molecules with respect to
CH$_3$OH (another, more abundant, grain mantle constituent) can 
be used to investigate the origin and evolution of these 
complex species
(e.g., Bottinelli et al. 2007). Using the abundance of CH$_3$OH
in B1 reported by BP97 (about $10^{-5}$), we derive
the abundance ratios HCOOCH$_3$/CH$_3$OH, HCOOH/CH$_3$OH, C$_2$H$_5$OH/CH$_3$OH,
and CH$_3$CN/CH$_3$OH to be in the order of $10^{-2}$, $10^{-3}$,
$10^{-3}$, 
and  $10^{-5}$ respectively. The first three are in agreement (within
an order of magnitude) with the
average abundance ratios found in hot cores and molecular clouds in
the GC region (Requena-Torres et al.~2006; Bisschop et al.~2007),
and are also consistent with the upper limits obtained toward
the L1448 outflow by Requena-Torres et al.~(2007).
Taken together these results suggest the dust in  
hot cores, GC molecular clouds, and the L1157 molecular cloud
have similar mantle composition
(as also argued by Requena-Torres et al.~2006; 2007).
In hot corinos, unlike the other sources, the abundance of HCOOCH$_3$ and HCOOH with respect to methanol is two orders of magnitude higher that in the B1 position of L1157. Such high relative abundance in hot corinos could be due to a longer warm-up phase of the gas surrounding low-mass protostars, when complex species are produced relatively more efficiently (Garrod \& Herbst 2006).
The CH$_3$CN to CH$_3$OH ratio is significantly lower 
by about three orders of magnitude
for B1 than
for hot cores and hot corinos 
(it has not been measured for GC clouds). 
One possible explanation for the
low CH$_3$CN/CH$_3$OH in B1 could be that 
processes in the shocked region
(but not present in hot cores or hot corinos) rapidly
destroy CH$_3$CN once it is in the gas phase. It is also
possible that CH$_3$CN is truly a daughter specie and it
shows very low abundance because there has not been
enough time ($< 2\times 10^3$ yr) for large abundances
of this molecule to form
in the warm gas associated with B1.
With our 
current data we cannot confidently state which
scenario is more likely and further
observations are needed. 

Similar abundance ratios of a number of molecules
in the L1157 outflow (and in hot cores) 
compared to the abundance ratio
of molecules in the comet Hale-Bopp led 
Bockel\'ee-Morvan et al.~(2000) to argue that there is
a direct link between cometary and interstellar ices. 
The estimates of HCOOCH$_3$/CH$_3$OH and HCOOH/CH$_3$OH
we obtain for L1157 are roughly similar to those of Hale-Bopp.
Our results are consistent with the conclusions reached by 
Bockel\'ee-Morvan et al.~(2000) that molecules in
cometary ices could have formed by processes very
similar to those that produce the chemically rich
icy mantles on interstellar grains. 


In summary, our results clearly indicate that in 
molecular clouds with only low-mass star formation 
complex organic molecules can form through 
grain surface reactions. Outflow shocks can
heat the surrounding medium and evaporate the icy
grain mantles where the complex species reside, thereby
releasing them into the gas phase and chemically
enriching the circumstellar environment. 
Our results show that a protostar's radiation
is not the sole mechanism that can 
generate complex molecules near forming stars
and that the impact of outflows needs to be considered when 
studying complex species around protostars.
If no subsequent chemical reactions alter the
abundance of complex molecules
once they are released into the gas phase by the shock,
then the abundance estimates from millimeter observations
can be used to study the mantle composition of the dust
in the cloud.  
Comparing the results from our observations
of the B1 clump in the L1157 outflow
with those of other regions observed by others 
suggest that
the grain mantle composition in the L1157 dark cloud
is comparable to that of the grains in hot cores and
molecular clouds in the Galactic center region. 
Observations of more complex molecules and estimates
of their relative abundance towards
other chemically active outflows will allow us
to determine the reliability of 
millimeter observations of shocked molecular gas 
for estimating 
the abundance of complex molecules in grains
and if similar grain mantle 
compositions are found in different regions of
low-mass star formation.


\newpage

\begin{deluxetable}{lcccccccc}
\tablecolumns{9}
\tabletypesize{\small}
\tablecaption{Detected Emission Lines
\label{detecttab}}
\tablehead{
\colhead{Molecule} & \colhead{Transition} &
\colhead{Frequency} & \colhead{$E_u/k$} & 
\colhead{$\delta v$\tablenotemark{a}} &
\colhead{$T_{mb}$\tablenotemark{b}} &
\colhead{$\Delta V$\tablenotemark{c}} & 
\colhead{$\int T_{mb} dV$\tablenotemark{d}} & \colhead{rms} \\
 &  \colhead{Line} & 
\colhead{[MHz]} &  \colhead{[K]} & 
\colhead{[km\ s$^{-1}$]} & \colhead{[mK]} &
 \colhead{[km\ s$^{-1}$]} &
\colhead{[K km\ s$^{-1}$]} &  \colhead{[mK]}
}
\startdata
HCOOCH$_3$-A   &  $7_{2,5}-6_{2,4}$   &  90156.48 &  19.68 & 2.1 & 11 & $5.0\pm1.5$  & $0.06\pm0.02$ &  2\\
               &  $8_{0,8}-7_{0,7}$   &  90229.63 &  20.07 & 1.0 & 12 & $3.7\pm1.5$ &  $0.05\pm0.02$ &  3\\
               &  $8_{3,6}-7_{3,5}$   &  98611.15 &  27.26 & 1.0 & 15 & $4.5\pm0.9$ &  $0.07\pm0.02$ &  3\\
               &  $8_{4,5}-7_{4,4}$   &  98682.60 &  31.90 & 3.0 &  7 & $3.3\pm2.5$ &  $0.03\pm0.03$ &  2\\
              & $20_{2,19}-19_{2,18}$ & 226718.70 & 120.27 & 0.8 & 26 & $2.3\pm0.8$ &  $0.07\pm0.03$ &  8\\

HCOOCH$_3$-E   &  $7_{2,5}-6_{2,4}$   &  90145.72 &  19.69 & 1.0 & 12 & $5.7\pm1.0$ &  $0.07\pm0.02$ &  2\\
               &  $8_{0,8}-7_{0,7}$   &  90227.63 &  20.10 & 1.0 & 12 & $3.0\pm0.7$ &  $0.04\pm0.01$ &  3\\
               &  $8_{3,6}-7_{3,5}$   &  98606.85 &  27.28 & 1.0 & 11 & $3.3\pm1.2$ &  $0.04\pm0.02$ &  3\\
               &  $8_{4,5}-7_{4,4}$   &  98712.06 &  31.91 & 3.0 &  7 & $5.5\pm2.3$ &  $0.04\pm0.02$ &  2\\
              
CH$_3$CN        & $14_{3}-13_{3}$  & 257482.80 &  156.77 & 1.2 & 26 & $4.7\pm1.2$ &    $0.13\pm0.04$ & 7\\
                & $14_{1}-13_{1}$  & 257522.50 &   99.89 & 1.2 & 23 & $4.0\pm1.0$ &    $0.10\pm0.03$ & 7\\ 
                & $14_{0}-13_{0}$  & 257527.40 &   92.78 & 1.5 & 34 & $3.6\pm1.0$ &    $0.14\pm0.04$ & 6\\ 

HCOOH          & $4_{2,2}-3_{2,1}$ &   90164.63 &  23.54 & 3.3 &  8 & $7.3\pm1.8$ &    $0.06\pm0.02$ &  2\\
               &  $6_{2,4}-5_{2,3}$ & 135737.76 &  35.48 & 1.4 & 17 & $5.8\pm1.4$ &    $0.11\pm0.03$ &  3\\

C$_2$H$_5$OH   &  $4_{1,4}-3_{0,3}$   &  90117.61 & 9.36 & 3.3 & 10 & $6.9\pm1.0$  &  $0.08\pm0.02$  & 2\\

\enddata  

\tablenotetext{a}{Width of velocity channel.}
\tablenotetext{b}{Peak intensity  from Gaussian fit to line.}
\tablenotetext{c}{Velocity width and error from Gaussian fit to emission line.}
\tablenotetext{d}{Errors in the integrated intensity estimate 
come from propagation of errors, using
the error in the velocity width and the rms.}

\end{deluxetable}

\begin{deluxetable}{lccc}
\tablecolumns{4}
\tabletypesize{\small}
\tablecaption{Estimates of Temperature, Column Density and Abundance
\label{rotdiagtab}}
\tablehead{
\colhead{Molecule} & 
\colhead{$T_{rot}$} & \colhead{$N$} & \colhead{$X=N/N_{H_2}$} \\
 & \colhead{[K]} & \colhead{[$10^{13}$ cm$^{-2}$]} & \colhead{[$10^{-8}$]} 
}
\startdata

HCOOCH$_3$-A\tablenotemark{a}   & $27\pm4$ & $15\pm4$ & $11\pm3$ \\

HCOOCH$_3$-E\tablenotemark{a}   & $18\pm13$ & $12\pm11$ & $8\pm7$ \\

CH$_3$CN\tablenotemark{a}   &  $110\pm50$  &  $0.1\pm0.05$ & $0.07\pm 0.04$\\

HCOOH\tablenotemark{b}    &  $10$ & $8$ & $5$ \\

C$_2$H$_5$OH\tablenotemark{c}  & 25\tablenotemark{d} & 10 & 7\\

\enddata
\tablenotetext{a}{Error estimates come from linear fit to rotational diagram.}
\tablenotetext{b}{No error estimate included as only two points were used for linear fit to rotational diagram.}
\tablenotetext{c}{No error estimate included as column density was not obtained using the rotational diagram, see text.}
\tablenotetext{d}{$T_{rot}$ assumed to be similar to the one derived
for HCOOCH$_3$.}

\end{deluxetable}

\clearpage


\begin{figure}
\plotone{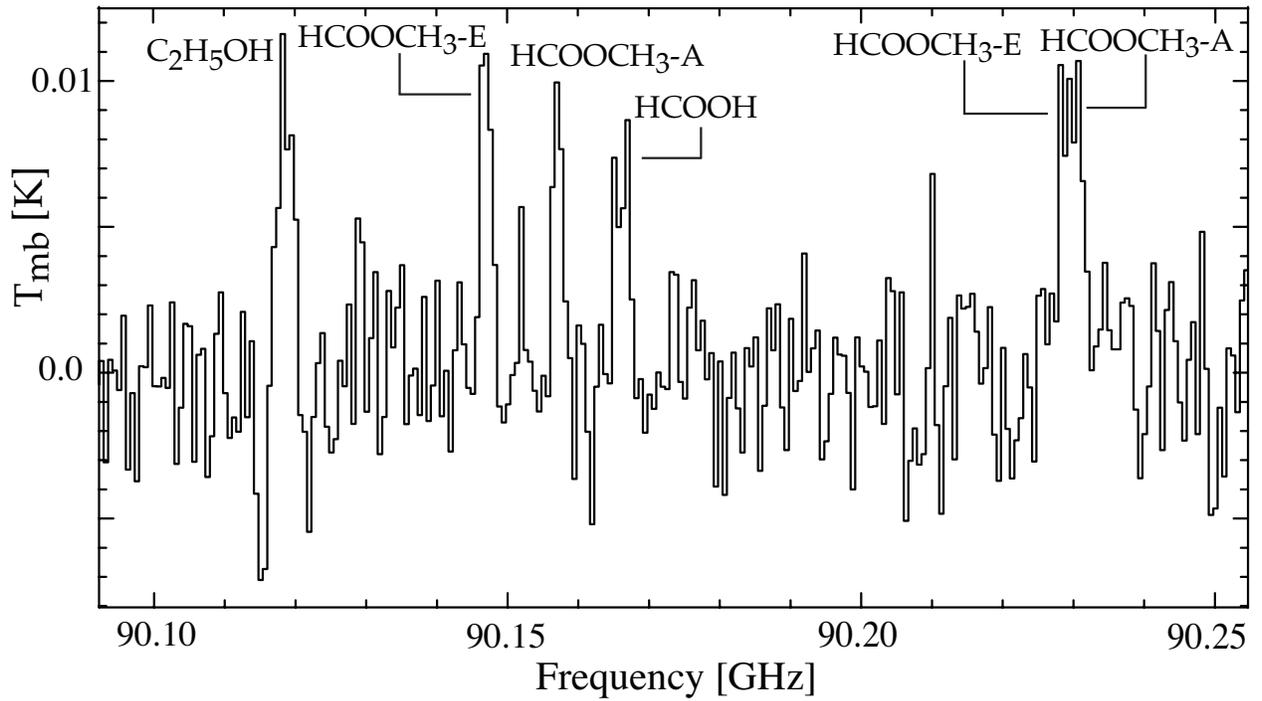}
 \caption{Sample spectrum from our observations of the B1 position in the L1157 outflow. 
The spectrum shown here is centered around 90.17 GHz, spans about 0.15 GHz, has a spectral
resolution of 0.625 MHz ($\sim 2$ km~s$^{-1}$) and an rms of 2 mK.
\label{specfig}}
\end{figure}

\clearpage

\begin{figure}
\plotone{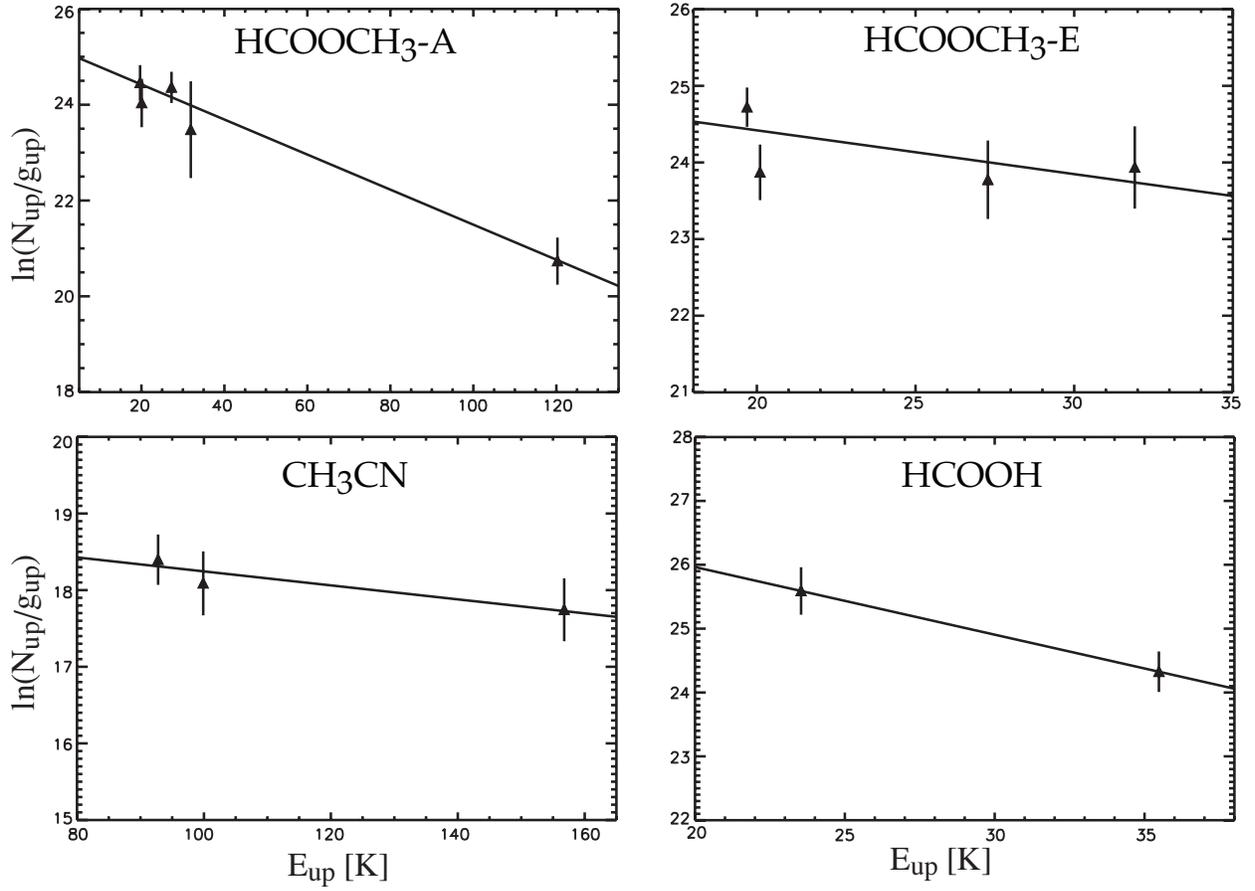}
 \caption{Rotational diagrams of molecules for which we detect two or more transitional lines.
Errors shown are derived from the errors in the integrated intensity (see Table 1). 
\label{rotdiagfig}}
\end{figure}

\end{document}